**A mismatch between self-efficacy and performance: Undergraduate women in engineering tend to have lower self-efficacy despite earning higher grades than men**

KYLE M. WHITCOMB[1], Z. YASEMIN KALENDER[1], TIMOTHY J. NOKES-MALACH[2], CHRISTIAN D. SCHUNN[2], and CHANDRALEKHA SINGH[1]

[1]Department of Physics and Astronomy, University of Pittsburgh, Allen Hall, 3941 O'Hara St, Pittsburgh, PA 15260, USA. E-mail: kmw136@pitt.edu, zyk2@pitt.edu, clsingh@pitt.edu

[2]Learning Research and Development Center, University of Pittsburgh, LRDC, 3939 O'Hara St, Pittsburgh, PA 15260, USA. E-mail: nokes@pitt.edu, schunn@pitt.edu

**Abstract**

There is a significant underrepresentation of women in many Science, Technology, Engineering, and Mathematics (STEM) majors and careers. Prior research has shown that self-efficacy can be a critical factor in student learning, and that there is a tendency for women to have lower self-efficacy than men in STEM disciplines. This study investigates gender differences in the relationship between engineering students' self-efficacy and course grades in foundational courses. By focusing on engineering students, we examined these gender differences simultaneously in four STEM disciplines (mathematics, engineering, physics, and chemistry) among the same population. Using survey data collected longitudinally at three time points and course grade data from five cohorts of engineering students at a large US-based research university, effect sizes of gender differences are calculated using Cohen's $d$ on two measures: responses to survey items on discipline-specific self-efficacy and course grades in all first-year foundational courses and second-year mathematics courses. In engineering, physics, and mathematics courses, we find sizeable discrepancies between self-efficacy and performance, with men appearing significantly more confident than women despite small or reverse direction differences in grades. In chemistry, women earn higher grades and have higher self-efficacy. The patterns are consistent across courses within each discipline. All self-efficacy gender differences close by the fourth year except physics self-efficacy. The disconnect between self-efficacy and course grades across subjects provides useful clues for targeted interventions to promote equitable learning environments. The most extreme disconnect occurs in physics and may help explain the severe underrepresentation of women in "physics-heavy" engineering disciplines, highlighting the importance of such interventions.

# 1. Introduction

Self-efficacy is intertwined with many aspects of academic life, so understanding both the nature of self-efficacy itself and its effects is crucial in order to improve education. Self-efficacy is the belief in one's capability to succeed in a particular task or subject [1-3], and in an academic context it can both affect and be affected by academic performance, a feedback loop that can either promote or hinder student learning [4-7]. Most importantly here, the feedback loop can produce growing inequities for those traditionally underrepresented in engineering [8-9]. Due to this recursive nature, measuring self-efficacy and academic performance longitudinally is vital to understanding their relationship. We present such a longitudinal analysis of the relationships between self-efficacy and academic performance using institutional data and survey responses from one US-based institution to test for patterns of gender differences. While the study is implemented within one US-based engineering program, the broader methodology described can be applied at any institution in order to seek sources of gender inequities inherent in their academic programs.

## 1.1 Underrepresentation of women in engineering fields

The overall underrepresentation of women in Science, Technology, Engineering, and Mathematics (STEM) careers is largely due to their large underrepresentation in engineering [10]. Progress has been made in some fields of engineering, but the largest engineering majors remain heavily male-dominated [10]. Many interrelated factors influence women's decision to pursue an education in engineering as well as subsequent decisions about which subfield to study and even whether to remain in engineering [11-15]. These factors include sociocultural factors, motivational factors, and various aspects of prior education such as quality of teaching [14-23]. In particular, it has been proposed that cultural bias and stereotypes can negatively impact the self-efficacy and academic performance of women in various STEM subjects such as mathematics and physics [15-18]. This is potentially damaging to prospective women in engineering since success in mathematics and science courses in high school has a positive impact on students' choice of and persistence in an engineering major [24-25] and longer-term career goals [5, 26].

## 1.2 Discipline-specific variation of gender differences in self-efficacy

Much of the research on self-efficacy has examined broad discipline self-efficacy (e.g., science or engineering self-efficacy) or more discipline-specific self-efficacy (e.g., physics or chemistry self-efficacy), especially as students become more specialized in tertiary education (e.g., become chemistry majors vs. physics majors).

Surprisingly, however, there has been little research comparing self-efficacy in various disciplines among the same population.

This gap in research comparing discipline-specific self-efficacy is particularly problematic for engineering. First, engineering students in all engineering majors take courses and apply content from many STEM disciplines, so self-efficacy in each discipline is relevant to them. Second, large differences in self-efficacy by gender have been previously found in various STEM disciplines [12-13, 21, 27-32] and have been implicated in gender differences in performance, retention, and choice of major [5, 24-26, 33], which is especially relevant during the students' first year. There are also hints that the size and even direction of self-efficacy differences vary by academic discipline. For example, women have equal or slightly higher self-efficacy in English [34] and equal self-efficacy in biology [35-36]. Third, it is difficult to generalize gender patterns in general-education STEM courses taken by many non-engineering majors to engineering students in particular because there are large variations in which students choose to take the general STEM courses (e.g., the population of students taking introductory biology is not the same as the population of students taking introductory physics) and because engineering students are potentially different from the general science major population (e.g., women choosing to major in the male-dominated engineering may have atypically high self-efficacy beliefs in physics and math relative to the general science major population). Fourth, women are often numerical minorities in the introductory courses because they are numerical minorities in engineering overall and therefore also in engineering-specific first year courses (e.g., General Chemistry for Engineers). Fifth, variation in gender differences in self-efficacy by discipline in later years might be explained by differential rates of participation in engineering majors by gender (e.g., male-dominated electrical and mechanical engineering vs. more balanced chemical engineering may reflect differential gender differences in physics vs. chemistry self-efficacy). Finally, due to the influence of biases and stereotypes about who can perform well in specific-disciplines, men and women may develop very different self-efficacy despite similar academic background [8-9].

Previous work has shown that women tend to have a lower self-efficacy than men in disciplines including physics [29-31] and chemistry [32] as well as mathematics and engineering [12-13, 21, 27-28]. However, these studies of physics, chemistry, and mathematics self-efficacy did not focus on engineering students.

**1.3 Time-varying gender differences in self-efficacy**

Another important gap in the self-efficacy research of particular relevance to engineering education is the lack of research on the change in student self-efficacy over time, particular across the full course of studies in engineering. Even if students successfully graduate with an engineering degree, they do not all enter the engineering workforce, and even those that do can exit shortly thereafter. The self-efficacy differences that were in place at the end of the degree will be more relevant to those later career transitions than self-efficacy that is typically measured within the first years of undergraduate studies. As course grades are generally higher in the more advanced courses [37] and experience with success accumulates, gender differences initially observed in the earlier years may disappear. On the other hand, due to the negative effects of having low self-efficacy on exam performance, the feedback loop from performance to self-efficacy, and the potentially negative effect of being a numerical minority in most courses (at least within some engineering majors), early gender differences could actually magnify over time.

A few studies have investigated self-efficacy changes over the course of two semesters, and generally found little change in self-efficacy over that period [31, 38-39]. One study found that over two years, the self-efficacy of women in engineering showed a positive trend [40]. Little is known about gender differences in self-efficacy of graduating engineering students.

**1.4 Gender differences in alignment of self-efficacy and academic performance**

There are multiple factors that lead to alignment of self-efficacy beliefs and academic performance. First, a central source of self-efficacy beliefs is prior performance feedback [4]. Second, self-efficacy can influence academic performance [4-7]. Third, factors that directly influence self-efficacy can also directly influence academic performance. In particular, stereotype threats that women experience in many STEM disciplines due to societal stereotypes and biases can increase their anxiety, rob them of cognitive resources while solving problems, and lead to reduced test scores [7]. Additionally, these stereotype threats can lower self-efficacy, which can result in reduced interest and engagement during learning [5, 41].

However, these factors may influence performance and self-efficacy in different manners, so academic success may not be fully aligned with self-efficacy. Of particular relevance here, alignment between academic performance and self-efficacy may be differential by gender. Women in STEM may interpret struggle in difficult courses as being due to inability whereas men may interpret struggle in these same courses as due to lack of effort [42]. Further, gendered stereotypes, differential availability of role models, and consistently being in a numerical minority in coursework may produce biased self-efficacy.

**1.5 Research Goals and Theoretical Framework**

The current paper is based on the idea that longitudinal measurements of student self-efficacy and academic performance in math, science, and engineering courses can provide invaluable insight into the long-term trends in these two intertwined aspects of undergraduate engineering students' characteristics that are the foundation for success and growth as an engineer.

Our theoretical framework draws upon the substantial prior research conducted on self-efficacy both as a general construct [4-7, 41] and as a discipline-specific construct [12-13, 21, 27-32]. Our work extends previous research by investigating trends within the same populations (engineering overall and by separate engineering majors) across four different STEM disciplines in order to situate any gender differences observed within the broader context of these students' academics. This research is of particular importance to engineering given the poor gender diversity in engineering overall and the negative role that low self-efficacy can play in students' academic decisions [11-15].

Studying self-efficacy gender differences and how they differ by STEM disciplines (e.g., math, physics, chemistry and engineering) for engineering students is important to better understanding gender differences in student choice of engineering major and future success in engineering careers, which have very large differences by gender [5, 24-26]. Furthermore, in a college or university setting, early interventions to improve male and female engineering students' self-efficacy in all relevant subjects may help ensure that all students excel. Otherwise, due to sociocultural stereotypes and biases pertaining to beliefs about who belongs in some engineering majors and is capable of excelling in associated foundational courses, e.g., in math and physics, students who do not fit the stereotypes will tend to have lower self-efficacy and perceive those majors as infeasible despite high interest and strong academic backgrounds.

The overarching goal of the research presented here is to spark a discussion amongst engineering educators and those teaching math and science courses for engineering students about strategies to create equitable and inclusive learning environments for all students regardless of their gender, particularly where there are gendered mismatches in self-efficacy and performance. This is of great importance for an interdisciplinary field such as engineering, where societal messages about any of the core subjects can have a pervasive impact on many aspects of students' academics. There is a particularly dangerous series of obstacles for women in engineering. First, being in a numerical minority can trigger anxiety due to stereotype threat during, for example, high stakes assessments. Second, along similar lines there are dangerous societal messages about only brilliant men being able to succeed in some fields (especially physics). Third, these sources of stereotype threat can reduce students' self-efficacy, which has been shown to have deep connections to academic performance and retention.

Our research questions to investigate the trends in self-efficacy and academic performance are as follows:

**RQ1.** Do men and women's self-efficacy in various core disciplines change along different trajectories as they progress from their first to their fourth year?

**RQ2.** Do gender differences in performance within foundational STEM courses vary by course discipline?

**RQ3.** Is there a match between gender differences in self-efficacy and gender differences in performance?

**RQ4.** Does being in a male-dominated engineering major change the gendered trajectories in self-efficacy relative to students in an more gender-balanced major?

## 2. Methodology

### 2.1 Participants

Using the Carnegie Classification of Institutions of Higher Education [43] , the US-based university at which this study was conducted is a public, high-research doctoral university, with balanced arts and sciences and professional schools, and a large, primarily residential undergraduate population that is full-time, more selective, and lower transfer-in. De-identified demographic data and university course grade data were provided by the university on all first-year engineering students who had enrolled from Fall 2009 through Spring 2018. We recognize that gender is not a binary construct; however, the data provided lists gender only as a binary categorical variable, so we present our analyses and results accordingly. Since all of our analyses will involve gender, we have filtered out students from the sample whose gender is unknown since they would later be omitted for each analysis.

The sample of engineering students for whom we have gender and grade data consists of 3,928 students. A subset of this sample also participated in surveys administered by the School of Engineering from Spring 2014 through Spring 2017 at the end of their first, second, and/or fourth years. The average response rate for each year was 79%. The full sample of students was 27% female and had the following race/ethnicities: 80% White, 8% Asian, 5% African American, 2% Latinx, and 5% Other. The mean age at the beginning of the students' first year was 18.9 years (SD=1.7 years), reflecting a population of students who predominantly are attending college immediately after completing high school.

**2.2 Measures**

*Grades.* The data provided include the grade points (GPs) earned in all courses at the university, the semester and class in which the course was taken, and the grade point distributions (mean and standard deviation) for each class. GPs are on a 0-4 scale (F = 0, D = 1, C = 2, B = 3, and A = 4) where '+' and '–' suffixes add or subtract 0.25 (e.g., B+ = 3.25) except for A+, which is recorded with a GP of 4.

*Declared major.* We also have the declared majors and minors for each student for every semester in which they were enrolled at the university. The structure of the School of Engineering at the university makes this a relatively accurate measure at the transition from students' first year to their second year since engineering students do not declare a major until the end of their first year when they choose an engineering major out of six possible choices. Therefore, for the purpose of measuring engineering students' intended majors from the end of their first year to the end of their second year, the first declared major is likely an accurate measure for most students. In particular, students can choose to major in Mechanical Engineering and Materials Science, Electrical and Computer Engineering, Chemical and Petroleum Engineering, Civil and Environmental Engineering, Bioengineering, or Industrial Engineering.

*Discipline-specific Self-efficacy.* The self-efficacy data were collected as part of an online survey that the engineering school gives to all engineering students at the end of the spring semester of their first, second, and fourth years. Students are given a few reminders to complete the survey and are told that this survey is important for evaluating the effectiveness of the engineering program, resulting in a completion rate exceeding 75% and sometimes higher than 90%. The four items used in this study consisted of responses to four prompts asking students to "Please rate your level of confidence in the following knowledge and skill areas: My ability to use my knowledge of [mathematics/engineering/physics/chemistry] to solve relevant engineering problems." The students were given five options – "poor," "fair," "good," "very good," and "excellent" – coded on a Likert scale from 1 to 5. Although some survey scales produce non-interval data that should not be analyzed as interval data, the Likert rating scales for measuring self-efficacy typically produce normally distributed data, as they did in the current study (see Appendix A). Further, these self-efficacy scales are always analyzed as interval data (e.g., by computing means, using t-tests and linear regressions).

It is considered best practice in the design of self-efficacy ratings to identify a particular task context to allow respondents to make reliable judgments [4]. The use of the phrase "to solve relevant engineering problems" in each survey question serves this purpose, in addition to increasing the relevance of the judgments to engineering education. However, it provides a small disconnect between the self-efficacy and performance contexts examined in this study: we examine performance in the core courses (e.g., performance in Calculus 1 and Physics 1), which rarely involve engineering problems *per se*. Doing well in a core course is not synonymous with being able to successfully apply that knowledge/ability to engineering problems. However, we argue that students very likely responded to these engineering-context self-efficacy judgments using general discipline-specific self-efficacy beliefs. In general, self-efficacy judgements for a science discipline that refer to different performance contexts (e.g., doing well in lab, doing well in tests, working on projects, understanding museum exhibits) tend to cohere as a single construct [31]. Further, a subset of students ($N = 446$) also completed a general physics self-efficacy survey within their Physics 2 class around the time that they were also completing the general engineering attitudinal survey, and the two measures were highly correlated ($r = 0.60$). Finally, the general physics self-efficacy survey showed a similar gender effect size (measured by Cohen's $d$) as the engineering-context physics self-efficacy ($d = 0.76$ in the physics context vs. $d = 0.84$ in the engineering context).

In smaller scale research studies, a survey scale often has multiple items so that scale reliability can be calculated (e.g., Cronbach alpha). In larger scale longitudinal studies, survey fatigue becomes a major concern as students stop responding with having to repeatedly do long surveys, and it is not uncommon to use only one survey item per scale [44-45]. In our case, the self-efficacy items were embedded in a larger survey and asking students three to five questions on self-efficacy per domain would have been received negatively. The main disadvantage of one-item scales is the increase in measurement noise (i.e., a problem of reliability, not a problem of validity), which reduces the ability to detect effects. This deficit is overcome by using a large sample, as in the current study. In a further effort to reduce survey fatigue, the survey is only administered to students at the end of their first, second, and fourth years. These time points capture the transition from the first to the second year as the students learn to adapt to their new academic environment as well as the longer-term transition from the early years to when the students ultimately earn their degrees.

**2.3 Analysis**

*Grouping by gender and major cluster.* For the entirety of this analysis, students are grouped by their reported gender in order to investigate gender differences in the perception of and performance in foundational subjects in engineering. To investigate whether these gender differences differ across different engineering departments, we used the first major declared by these engineering students, which typically occurs at the end of their first year when the students move from the standard first-year courses to a specific engineering department's curriculum.

Students were grouped into three clusters of engineering majors determined by the proportion of women in those majors. Typically, these engineering students first declare their major at the end of their first year. The specific majors that went into each cluster were determined by the proportions of women and men who declared that major. Table 1 shows the engineering majors in each major cluster as well as the number of students and percentage of that number that are women, both for the full sample and the sample of survey takers.

Table 1: The sample size of students in each major cluster, along with the percentage of women in that sample. Two samples are reported for each cluster, one for all students for which we have grade data and another for the subset for which we have survey data. Survey takers counted here may have taken any combination of the first year, second year, and fourth year surveys.

| Cluster | Disciplines | All Students | | Survey Takers | |
|---|---|---|---|---|---|
| | | N | % Women | N | % Women |
| 1 | Electrical Computer Mechanical | 2128 | 16% | 991 | 22% |
| 2 | Chemical Environmental Civil | 1551 | 34% | 738 | 37% |
| 3 | Bioengineering Industrial | 1185 | 41% | 700 | 46% |

*Self-efficacy differences by gender.* In order to test for statistically significant differences in self-efficacy scores, we performed $t$-tests comparing self-efficacy scores of men and women in engineering. Effect sizes of the gender differences were calculated in standard deviation units via Cohen's $d$ [46]. Analyses of gender differences in self-efficacy were run separately for each of the four disciplines (mathematics, engineering, physics, and chemistry) at each time point (but averaging across the five cohorts of students). Furthermore, these tests were run for all available students first, then separately for the men and women in each major cluster.

*Course performance differences by gender.* Similarly, using $t$-tests [47-48], we investigated gender differences in course performance on the grade points earned by men and women in each of the foundational courses.

Again, the magnitudes of differences were calculated in standard deviation units. The courses we investigated were the foundational courses taken by the largest number of students in the School of Engineering, namely all of the common first-year courses in engineering, physics, chemistry, and mathematics as well as a selection of second-year mathematics courses taken by students in a variety of engineering departments: two in chemistry, two in physics, two in engineering, and five in mathematics. We note that the introductory engineering sequence at the studied university is a two-course sequence designed to teach the students computer programming skills in an engineering context, and in particular teaches the students to use MATLAB, C++, and Python to solve engineering problems. Further, we recognize that the content of such a course, or of the introductory curriculum as a whole, may vary from country to country and even institution to institution within a country. Thus, the investigation presented here provides a methodology for investigating these relationships within any particular curriculum, and the results presented here may be used as a comparison to a particular institution with a strictly enforced first-year engineering curriculum that includes courses in chemistry, engineering, mathematics, and physics.

## 3. Results and Discussion

### 3.1 Longitudinal Gender Differences in Self-Efficacy

In order to understand the perceptions of these engineering students about their foundational course work and answer Research Question 1, and specifically address how these perceptions differ for men and women, we plot in Figure 1 the mean self-efficacy scores of men and women in engineering in each of the four foundational subjects (mathematics, engineering, physics, and chemistry) at each time point (end of the first, second, and fourth years). The full distributions of responses to these prompts are available in Appendix A. Looking at the data for the first year, there is a statistically significant gender gap favoring men in self-efficacy scores for applying mathematics, engineering, and physics to their work in engineering, and no difference in chemistry. Both mathematics and engineering follow a similar trajectory in that the initial gap remains in the second year and is eliminated by the fourth year. In sharp contrast, the gap shrinks but remains relatively large in physics even by the end of the fourth year. At no point is there a significant gender difference in chemistry self-efficacy.

Although we do not focus on this issue here, we note that self-efficacy of both men and women appears to grow, as expected, over years; the lack of growth in chemistry self-efficacy may reflect the relatively small role chemistry plays in the largest majors within the six engineering departments. Note that there is relatively little change in majors after students declare majors in their second years, nor is there much attrition overall in Engineering at this university after the second year. Thus, the changes between second and fourth years are unlikely to be caused by major switching or attrition. These arguments are further supported by follow-up analyses that included only data from students completed surveys at all three time points.

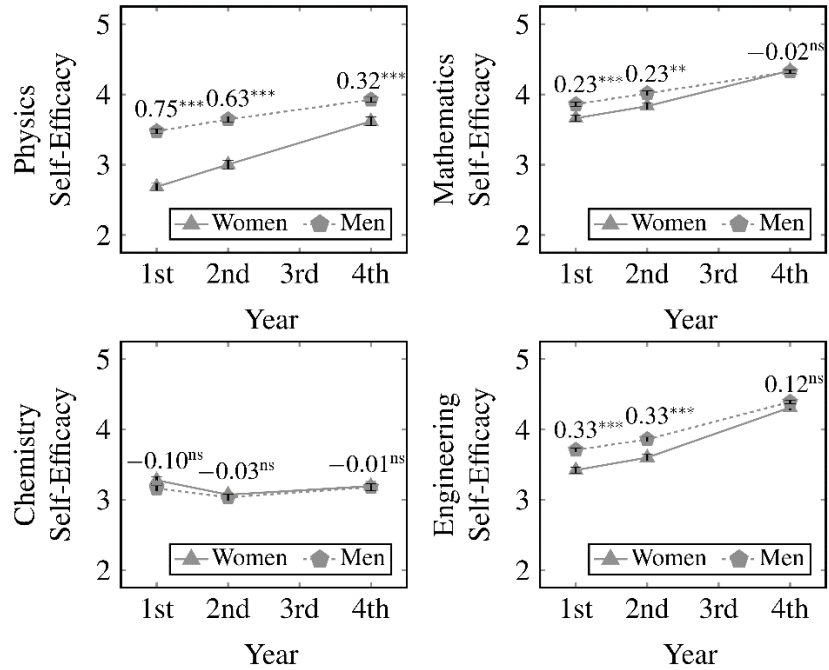

Figure 1: The mean self-efficacy scores of engineering students at the end of their first, second, and fourth years in each of the foundational subjects in engineering are plotted along with their standard error. Self-efficacy was measured on a Likert scale from 1 to 5. The vertical range of self-efficacy scores has been restricted to better show the gender differences. Above each pair of points, Cohen's $\boldsymbol{d}$ is reported (with $\boldsymbol{d} < \boldsymbol{0}$ indicating women have a higher mean and $\boldsymbol{d} > \boldsymbol{0}$ indicating men have a higher mean) along with the statistical significance of the gender difference according to a $\boldsymbol{t}$-test, with $^{*}\boldsymbol{p} < \boldsymbol{0.05}$, $^{**}\boldsymbol{p} < \boldsymbol{0.01}$, $^{***}\boldsymbol{p} < \boldsymbol{0.001}$, and $^{\text{ns}}\boldsymbol{p} > \boldsymbol{0.05}$.

### 3.2 Performance Differences in Foundational Courses

Gender differences in performance were investigated across the foundational first year engineering curriculum and selected common second-year mathematics courses to answer Research Question 2. Table 2 reports the summary statistics (population $N$, mean $\mu$, and standard deviation $\sigma$) for each course along with a $p$-value from a $t$-test comparing the grades earned by men and women in that course and the effect size (Cohen's $d$).

Table 2: Reported are the performance differences between female (F) and male (M) engineering students for grades earned in introductory courses in engineering, physics, chemistry, and mathematics as well as second-year mathematics courses. We report the sample size $N$, mean $M$, and standard deviation $SD$ separated for men and women, as well as the $p$-value from a $t$-test comparing the grades earned and Cohen's $d$ measuring the effect size, both for the individual courses and for each of the four subjects overall. The sign convention for Cohen's $d$ matches that of Figure 1.

| Course | Gender | $N$ | $M$ | $SD$ | $p$ | Cohen's $d$ Course | Cohen's $d$ Subject |
|---|---|---|---|---|---|---|---|
| Engineering 1 | F | 1070 | 3.68 | 0.38 | < 0.001 | -0.19 | -0.12 |
|  | M | 2715 | 3.60 | 0.44 |  |  |  |
| Engineering 2 | F | 1095 | 3.33 | 0.66 | 0.168 | -0.05 |  |
|  | M | 2797 | 3.30 | 0.72 |  |  |  |
| Physics 1 | F | 1127 | 2.60 | 0.82 | < 0.001 | 0.19 | 0.14 |
|  | M | 2824 | 2.75 | 0.86 |  |  |  |
| Physics 2 | F | 1123 | 2.59 | 0.85 | 0.015 | 0.08 |  |
|  | M | 2998 | 2.67 | 0.91 |  |  |  |
| Chemistry 1 | F | 1062 | 2.91 | 0.86 | < 0.001 | -0.13 | -0.15 |
|  | M | 2688 | 2.80 | 0.91 |  |  |  |
| Chemistry 2 | F | 1010 | 2.79 | 0.85 | < 0.001 | -0.17 |  |
|  | M | 2514 | 2.64 | 0.91 |  |  |  |
| Calculus 1 | F | 857 | 3.04 | 0.90 | 0.002 | -0.12 | -0.15 |
|  | M | 2199 | 2.92 | 0.96 |  |  |  |
| Calculus 2 | F | 989 | 2.87 | 1.00 | < 0.001 | -0.12 |  |
|  | M | 2563 | 2.74 | 1.08 |  |  |  |
| Calculus 3 | F | 1193 | 2.88 | 1.03 | < 0.001 | -0.12 |  |
|  | M | 2954 | 2.75 | 1.10 |  |  |  |
| Linear Algebra | F | 762 | 3.30 | 0.91 | < 0.001 | -0.25 |  |
|  | M | 2296 | 3.05 | 1.06 |  |  |  |
| Differential Equations | F | 1176 | 2.89 | 1.04 | < 0.001 | -0.14 |  |
|  | M | 3189 | 2.74 | 1.12 |  |  |  |

For all but one course, there were statistically significant gender differences, with all but one of those statistically significant results satisfying $p < 0.01$. Most interestingly, the direction of the differences varied by discipline. Only for the two introductory physics courses did men receive higher grades on average than women. In all the courses in the other three disciplines, women received higher grades on average than did men (and statistically significantly so except for Engineering 2). Moreover, although the lowest mean grades occurred in physics, the gender patterns in physics cannot be explained by physics being the most difficult course because students had similarly low grades in chemistry and calculus but with opposite gender differences (with women on average performing better than men in all chemistry and mathematics courses). Similarly, the differences could not be explained in terms of the stronger role of mathematics in physics versus chemistry because women had higher grades in every single mathematics course.

It should be acknowledged, however, that none of the gender differences in course performance was large. Instead, what is surprising is the pattern of large gender differences in self-efficacy despite small differences in performance as well as performance and self-efficacy scores showing opposite trends (e.g., in mathematics and engineering, women on average have better grades but have lower self-efficacy than men). This contrast is directly examined in the next section.

### 3.3 The Relationship Between Self-Efficacy and Course Performance

In order to answer Research Question 3 and investigate the relationship between self-efficacy and performance, we combined the two previous analyses for Research Questions 1 and 2 to plot the effect sizes of gender differences in both self-efficacy and course grades (Figure 2). For each point in the plot, we used only the subsample for which we had both a course grade (Ns varying from 579 for Linear Algebra to 1,163 for Engineering 1) and a self-efficacy score in the nearest survey (first year for the introductory courses, and second year for the second-year mathematics courses), a restriction which slightly alters the effect sizes from those of the full samples in Figure 1 and Table 2. In addition to dashed lines along $d = 0$ on both axes, there is a dotted line along $d_{SE} = d_{CG}$ (where the effect size of self-efficacy differences equals the effect size of course grade differences), which represents where the data would fall if there was a one-to-one relationship between the effect sizes of self-efficacy and course grade. In addition, a vertical line is shown from the center of each discipline to the $d_{SE} = d_{CG}$ line, which represents the deviation of self-efficacy differences from academic performance differences.

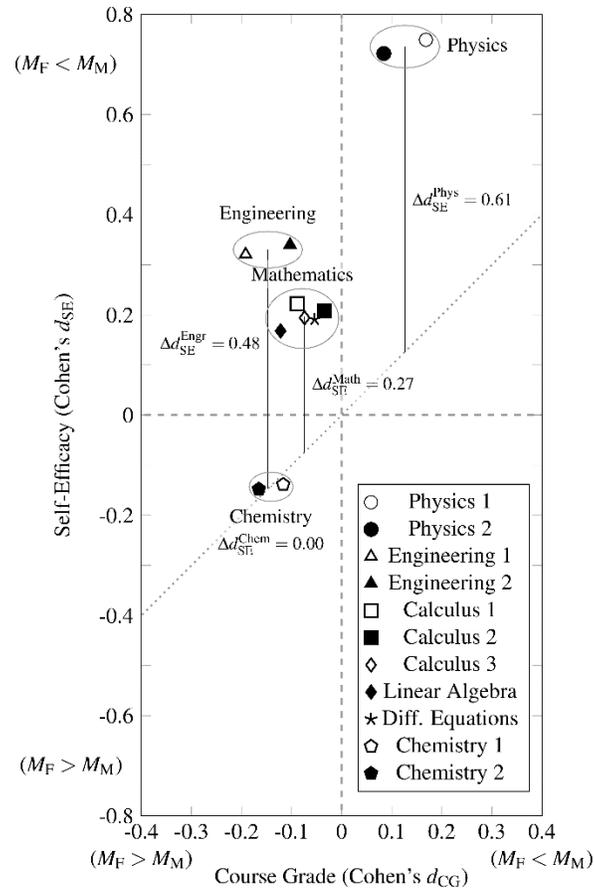

Figure 2: The effect sizes (Cohen's $d$, sign convention matching Figure 1) of gender differences in self-efficacy ($d_{SE}$) and course grades ($d_{CG}$) are plotted for each of the introductory courses as well as second-year mathematics courses. Dashed lines for $d = 0$ on both axes as well as a dotted line along $d_{SE} = d_{CG}$ have been added. Ellipses group all courses in each subject. Each point contains the data of only those students for which both a grade and self-efficacy score were available. Introductory course grades (the first and second courses in a sequence) are paired with first-year self-efficacy scores, while second-year mathematics course grades are paired with second-year self-efficacy scores. Vertical lines have been added showing the distance along the self-efficacy axis from the center of each subject (defined as the average position of the constituent courses) to the $d_{SE} = d_{CG}$ line.

Here we see the trends among the foundational subjects shine through strongly. As before, we see that engineering and mathematics are more similar than the other two disciplines. However, it is highly noteworthy that the directions of the self-efficacy and performance gender differences for all courses within two subjects are in direct opposition. In both engineering and mathematics, men have a higher self-efficacy, while women earn higher grades on average.

In Figure 1, chemistry was the only subject in which there was no gender difference in self-efficacy, though we did see a performance difference favoring women. In Figure 2, with the subset of the population for which we have both self-efficacy scores and course grade data, we see that chemistry falls right along the $d_{SE} = d_{CG}$ line. That is, for the populations completing the surveys, there is direct correspondence between self-efficacy and performance: women have higher self-efficacy in chemistry to the exact same extent to which they also outperform men in chemistry.

Figure 2 also shows an interesting contrast between quadrant and deviation from the matching self-efficacy/performance line. Both chemistry and physics are in 'match' quadrants of the figure, but while chemistry is precisely on the $d_{\text{SE}} = d_{\text{CG}}$ line, physics is not. On the one hand, men have higher self-efficacy in applying physics to engineering and do have higher performance, which is roughly a match. On the other hand, effect sizes are completely mismatched: the self-efficacy gap is much larger than the course grade gap leading to physics lying further away from the $d_{\text{SE}} = d_{\text{CG}}$ line than any other discipline.

**3.4 Self-Efficacy Time Trends for Different Major Clusters**

In Figure 3, as in Figure 1, we plot the mean self-efficacy scores of men and women in each of the four foundational subjects, but now separately for each major cluster. The three clusters of majors considered are 1) mechanical/materials science and electrical/computer engineering; 2) chemical and environmental/civil engineering; and 3) bioengineering and industrial engineering (see Table 1). The mean responses in each subject for each of the six majors separately are in Appendix B. Looking first at mathematics and engineering, we observe some small differences across the major groups, namely clusters 1 and 2 have self-efficacy differences in the first and second years that close by senior year while cluster 3 shows no self-efficacy differences in mathematics or engineering except a marginally significant gap in mathematics self-efficacy in the fourth year. As before there is a high degree of similarity between the responses in relation to mathematics and engineering within each cluster, suggesting that the strong tie in engineering students' perceptions of their abilities in mathematics and engineering is true for all engineering students, even despite overall differences between cluster 3 and clusters 1 and 2.

The third row of Figure 3, showing responses for physics self-efficacy, shows a higher physics self-efficacy among students who choose physics-oriented majors such as electrical and mechanical engineering than students in other majors. Most importantly, Figure 3 sheds additional light on likely causes of the unusually large and not closed gender gap in physics self-efficacy across majors. The cluster with the highest percentage of women (bioengineering/industrial engineering) is actually the only subset where the physics self-efficacy gap remains open by the fourth year. In other words, being in courses that are male dominant (at least by numbers) does not appear to be the central cause of the effect.

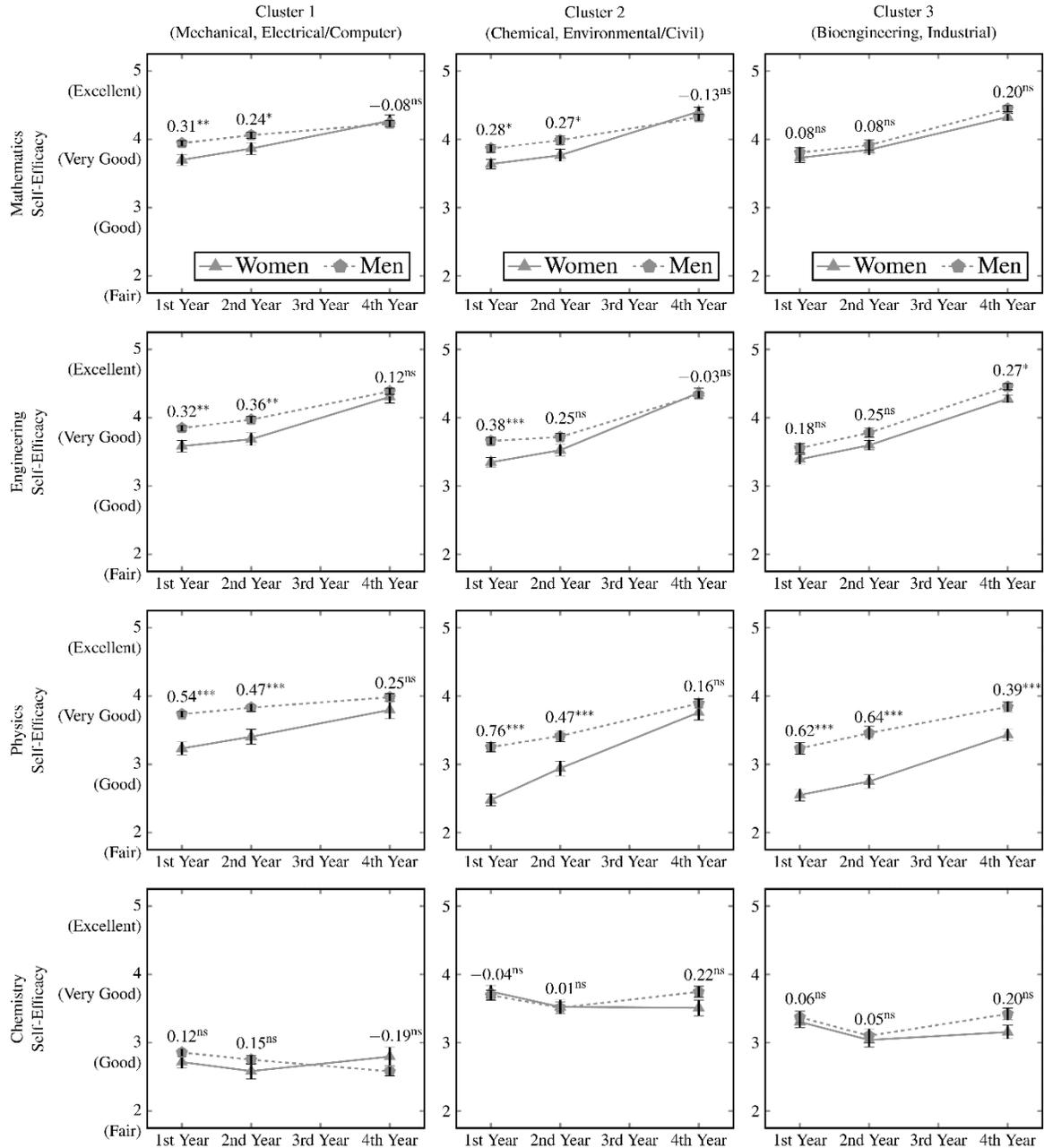

Figure 3: As in Figure 1, the mean self-efficacy scores of engineering students at the end of their first, second, and fourth years in each of the foundational subjects in engineering are plotted along with their standard error separately for students in each of the three clusters of majors. Self-efficacy was measured on a Likert scale from 1 to 5. The range of self-efficacy scores has been restricted to better show the gender differences. Each column contains the graphs for the different major groups while each row contains the graphs for self-efficacy in the different foundational subjects. Above each pair of points, effect size (Cohen's $d$, sign convention matching Figure 1) is reported along with the statistical significance of the gender difference according to a $t$-test, with $^*p < 0.05$, $^{**}p < 0.01$, $^{***}p < 0.001$, and $^{ns}p > 0.05$.

A related point has to do with ruling out simple exposure to physics content. If the gender effect in physics were simply a result of a relative exposure to physics content, then we should expect that men in cluster 3 also have lower physics self-efficacy than students in cluster 1. While that is somewhat true for the first two years – possibly driven by a selection effect with the students that have the highest physics self-efficacy being most likely to choose a mechanical, electrical, or computer engineering major – that gap closes for men in cluster 3 by the fourth year, but not for women in the same majors. Further, it is noteworthy that the women in the physics-oriented majors in cluster 1 display a remarkably similar trend in physics self-efficacy to the men in clusters 2 and 3.

Finally, the fourth row of Figure 3 shows the chemistry self-efficacy scores for each of the major groups, again showing a disparity based on chosen major. Cluster 2, which includes chemical engineering, has the highest chemistry self-efficacy across all years, followed by cluster 3, and finally cluster 1, students in which seem to display a decrease in their perceived to applying chemistry to engineering that persists through the fourth year. Despite having no gender gaps, Figure 3 does show strong differences between the three major clusters, consistent with the idea that self-efficacy may play an important role in major selection (i.e., the highest self-efficacies occur from the start in students who selected chemical engineering). Turning to lower exposure/experience effects, Cluster 1's chemistry self-efficacy scores are the only ones which remain below 3 on the Likert scale even through their fourth year. Interestingly, a very small (albeit non-significant) gender gap in chemistry self-efficacy emerges in all groups by the fourth year, with men on average having a slightly higher chemistry self-efficacy than women in clusters 2 and 3 (including chemical engineering and bioengineering) and women on average having a higher chemistry self-efficacy than men in cluster 1. However, none of these differences by gender or time in chemistry self-efficacy were large.

## 4. General Discussion

Research Question 1 considers gender differences and temporal persistence of gender differences in engineering students' self-efficacy within four core disciplines. We observed no significant gender differences in chemistry self-efficacy in this undergraduate engineering population while that same population showed consistent differences in mathematics, engineering, and physics, with men tending to have a higher average self-efficacy in the first two years and the gap reducing by the fourth year. These effects were very small in mathematics and engineering, but large in physics. Given that a shared population of students was investigated (and they were all engineering students), this was the first clue that that are discipline-specific biases at play underlying gender differences in self-efficacy.

Turning to RQ2, we also observed consistent gender differences in course grades in each of the four core disciplines. Here however, women performed slightly better than men on average in mathematics, engineering, and chemistry. Such course differences were consistent with trends within this population for high school GPA differences by gender, and somewhat consistent with national trends [49], where mathematics shows mixed gender differences across institutions, although the same study finds men earn higher grades on average in chemistry. In contrast, men performed slightly better on average than did women in the two physics core courses that were examined. Thus, both self-efficacy and performance display substantial variation in gender differences by core STEM discipline, but not in consistent ways.

Focusing more specifically on alignment of self-efficacy with performance for RQ3, we observed three different patterns: 1) complete alignment with chemistry; 2) opposing direction small effects in mathematics and engineering; and 3) consistent direction but mismatching effect sizes in Physics (i.e., large self-efficacy effects but small performance differences). Thus, in answer to RQ3, gender differences in self-efficacy are inconsistently aligned with performance differences, strongly suggestive of some source of bias by gender in self-efficacy estimates that is discipline specific.

RQ4 investigated one possible source of such biases in self-efficacy: the experience of being a numerical minority in course work [8-9]. Here the self-efficacy trends through the later years was most relevant because it is via those courses that the experience of engineering students varies by major. However, the data did not support an effect of being a numerical minority: the change patterns were generally similar across majors. The one exception actually went in the opposite direction: the large physics self-efficacy gender difference was primarily found in the majors with the highest proportion of women.

While these analyses are inherently correlational, the correlational pattern has ruled out some commonly offered explanations for gender differences in performance and self-efficacy. First, the self-efficacy gender differences cannot be a simple reflection of actual performance differences; if anything, the pattern is more consistent with self-efficacy biases causing performance differences. Second, performance differences in physics cannot be attributed to deficits in mathematical ability: among engineering students, women outperform men on average in every single mathematics course. Third, self-efficacy differences by gender do not seem to be driven by the experience of being a numerical minority in coursework.

**4.1 Implications for Instruction and Future Research**

The observed correlational findings have important implications for research and practice. We divide those implications around the sources of initial differences by discipline and then the differential change over the years by discipline. Focusing first on the initial differences, other work [29-31] has shown that engineers show large physics self-efficacy differences by gender even in the first few weeks of class in their first year. The partial persistence of the differences throughout the degree argues that it is important to better understand the location of the effect within K-12 schooling and broader society sources. Physics tends to involve particularly strong beliefs that talent is required for doing well and the common cultural stereotype is that males are the ones that have such talent [50-52]. But it is unknown whether these effects happen through messages in the home, the media, from friends, or from science teachers, especially for engineering students who presumably had some relatively positive experiences that led them to pursue engineering. To address these sources directly, the ones having the largest effects on this population must be identified.

Regardless of the source, counter programming in secondary schools should be introduced to broadly address stereotypes and improve women's self-efficacy for physics (e.g., using growth mindset interventions [53-55]). If broadly distributed, the number of women applicants to engineering schools could rise. If more narrowly targeted to incoming engineering first-year students (a task more in the control of universities), prior biases might be lessened just in time to reduce stereotype-threat effects within the early coursework. Because the self-efficacy difference is particularly large in physics, physics-specific messages could be a good pragmatic choice, and such interventions could shift the large gender imbalance of students selecting physics-heavy engineering majors.

Turning to the differential changes in self-efficacy during the undergraduate years, at least this particular institution could perhaps be congratulated for having been able to reduce and sometimes entirely eliminate initial gender differences in self-efficacy and generally support a growing set of competence in the students. Future research should examine how broadly engineering programs around the world are able to achieve this outcome. Retention patterns at this university were not unusual by national US trends [10], which suggests that many US-based universities have also been similarly successful.

However, there are still important challenges: differences in self-efficacy remained overall, particularly in bioengineering and industrial engineering majors. Additional research should now focus on the experience of these students to understand why their physics self-efficacy remains so low. Physics is an important foundation to bioengineering overall and to many aspects of industrial engineering. Along these lines, replications at other universities are needed to examine whether these patterns are characteristics of those majors more generally or whether they come from department-specific messaging and coursework at this US-based university. Multiple cohorts were examined to produce sufficient power and allow for a range of instructors to have played a role, but nonetheless there can be institution-specific factors.

The current study was limited in that it did not follow students beyond the undergraduate experience. It is therefore not known what long-term consequences of the physics self-efficacy gap is. For example, does the gap further decrease with more real-world engineering experience, or does it further drive the gendered attrition within engineering that has been documented at the later career stages.

Another gap to acknowledge in the current study is the exclusive focus on gender and gender as a binary construct. Engineering is not only male-dominated: there is also underrepresentation by other gender identities and by sexual orientation [56-59], by race/ethnicity, and there can be additional effects at the intersection of race/ethnicity and gender (e.g., larger gender effects within underrepresented minorities). The current study offers an analytic framework that could equally well be applied to better understand those other domains of underrepresentation, even if they potentially have different explanations.

## 5. Conclusions

Self-efficacy is a specific attitude that can play a strong role in influencing student performance in both the short-run (by undermining studying and exam performance) and in the long-run (by influencing degree persistence). In this study, we add important nuances to the common previous finding of lower self-efficacy by women in STEM. In particular, we show that the self-efficacy gender difference varies by specific STEM topic, even within a particular sample of engineering students. Further, we also draw attention to the non-normativity of these self-efficacy gender differences: women sometimes have lower self-efficacy even when they have higher performance. On a more positive note, we observed reductions in the self-efficacy gap over time, but the early gender gaps are still important because they can affect differential attrition, which is generally highest in the early years of the engineering degree. Finally, we show that although there is some variation in these patterns by engineering major, there is no support for larger gender-based self-efficacy gaps being found in majors where women are a numerical minority.

**Acknowledgments**

This research is supported by NSF Grant DUE-1524575, Alfred P. Sloan Foundation Grant No. G-2018-11183, and the James S. McDonnell Foundation.

**Biography**

**Kyle Whitcomb** is a Ph.D. candidate in the Department of Physics and Astronomy at the University of Pittsburgh. He obtained his B.S. in Physics and Mathematics from the University of Puget Sound. His research focuses on promoting equity and inclusion in undergraduate physics education and using statistical methods to identify sources of inequities courses and how these inequities propagate through students' academic careers.

**Z. Yasemin Kalender** obtained a B.S. degree from Boğaziçi University in Istanbul with a focus on experimental particle physics during which she worked as a collaborator in several projects at CERN. Later, she moved to U.S. to continue her education as a Ph.D. student at the University of Pittsburgh with a research focus on physics education. Her current research interest is investigating motivational characteristics of students, diversity issues in the physics discipline, and incorporating big data analysis techniques into the area of physics education research. After earning her Ph.D. in 2019, she continues her academic training as a postdoctoral research associate at Cornell University.

**Timothy Nokes-Malach** is an Associate Professor of Psychology and a Research Scientist at the Learning Research and Development Center at the University of Pittsburgh. He received his Bachelors degree from the University of Wisconsin-Whitewater, PhD from the University of Illinois at Chicago, and Postdoctoral training at the Beckman Institute for Advanced Science and Technology at the University of Illinois at Urbana-Champaign. His research focuses on human learning, motivation, and knowledge transfer. His work has been supported with grants from the James S. McDonnell Foundation, the Pittsburgh Science of Learning Center, the National Science Foundation, and the Department of Education's Institute for Education Sciences.

**Christian Schunn** obtained his PhD from Carnegie Mellon in 1995. He currently is a Senior Scientist at the Learning Research and Development Center and a Professor of Psychology, Learning Sciences and Policy, and Intelligent Systems at the University of Pittsburgh. Most recently he became Co-Director of the Institute for Learning. He has led many research and design projects in science, mathematics, engineering, technology, and writing education. His current research interests include STEM reasoning (particularly studying practicing scientists and engineers) and learning (developing and studying integrations of science & engineering or science & math), neuroscience of complex learning (in science and math), peer interaction and instruction (especially for writing instruction), and engagement and learning (especially in science). He is a Fellow of several scientific societies (AAAS, APA, APS) as well as a Fellow and Executive member of the International Society for Design & Development in Education. He has served on two National Academy of Engineering committees, K12 Engineering Education and K-12 Engineering Education Standards.

**Chandralekha Singh** is a professor in the Department of Physics and Astronomy and the Director of the Discipline-based Science Education Research Center at the University of Pittsburgh. She obtained her Ph.D. in theoretical condensed matter physics from the University of California Santa Barbara and was a postdoctoral fellow at the University of Illinois Urbana Champaign, before joining the University of Pittsburgh. She has been conducting research in physics education for more than two decades. She was elected to the Presidential-line of the American Association of Physics Teachers and is currently serving as the President-Elect. She held the Chair-line of the American Physical Society Forum on Education from 2009-2013 and was the chair of the editorial board of Physical Review Special Topics Physics Education Research from 2010-2013. She was the co-organizer of the first and third conferences on graduate education in physics and chaired the second conference on graduate education in physics in 2013. She is a Fellow of the American Physical Society, American Association of Physics Teachers and the American Association for the Advancement of Science.

# APPENDIX A: Self-Efficacy Distributions

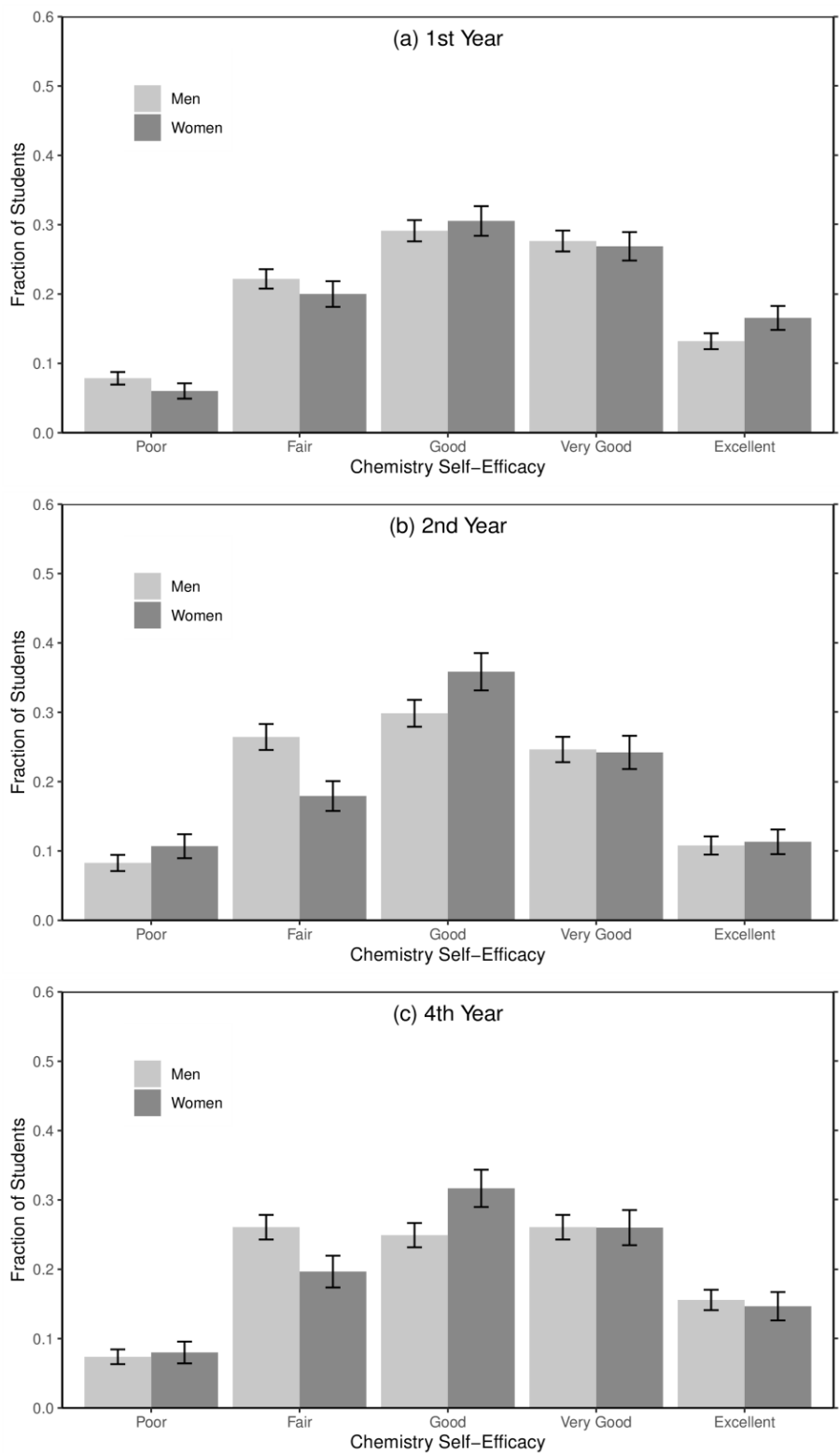

Figure 4: Distribution of responses to the chemistry self-efficacy prompts the surveys taken by students in their (a)1st year, (b) 2nd year, and (c) 3rd year. The fraction of men and women, respectively, who answered each of the five options are plotted along with the standard error.

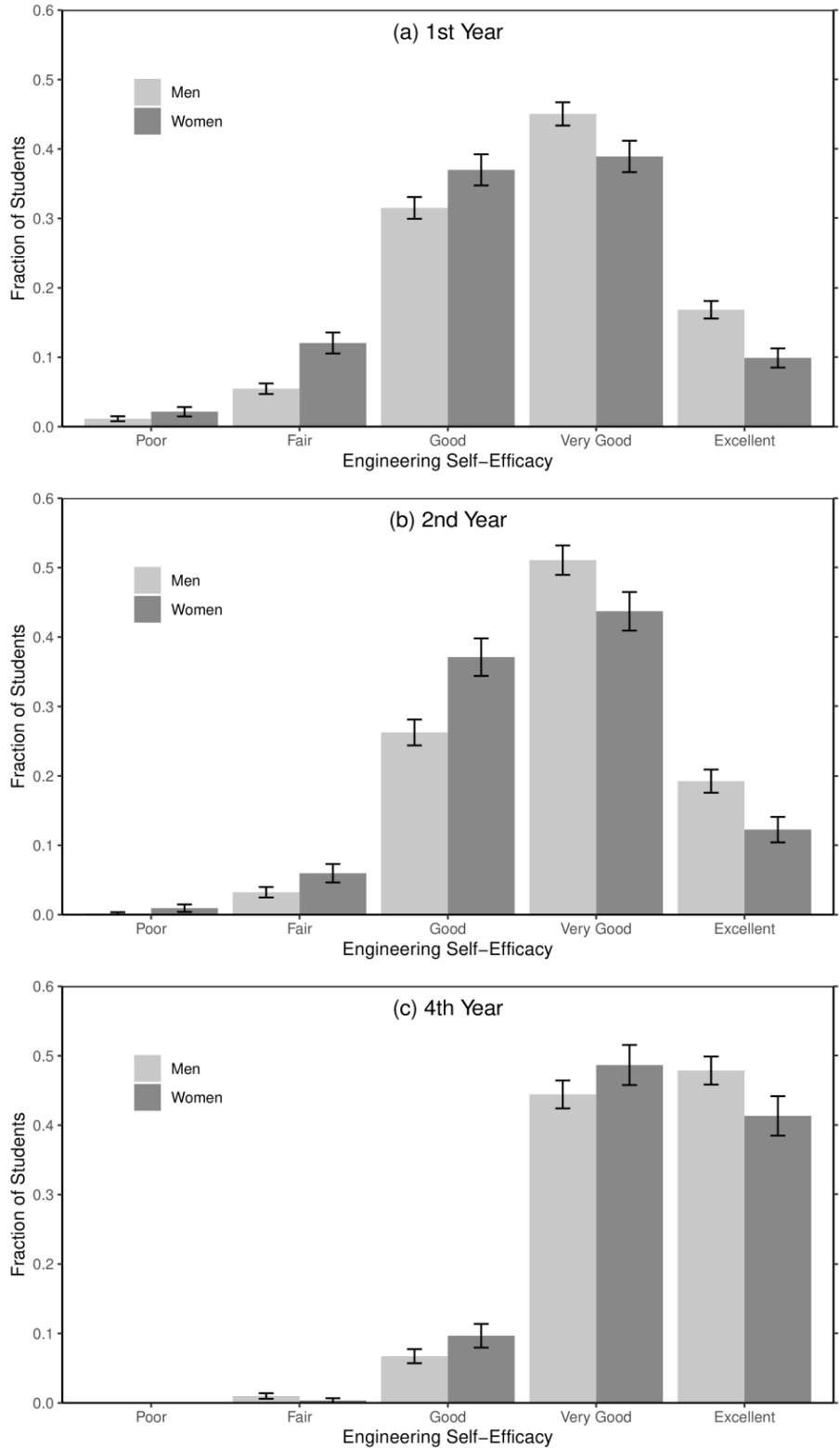

Figure 5: Distribution of responses to the engineering self-efficacy prompts the surveys taken by students in their (a) 1st year, (b) 2nd year, and (c) 3rd year. The fraction of men and women, respectively, who answered each of the five options are plotted along with the standard error.

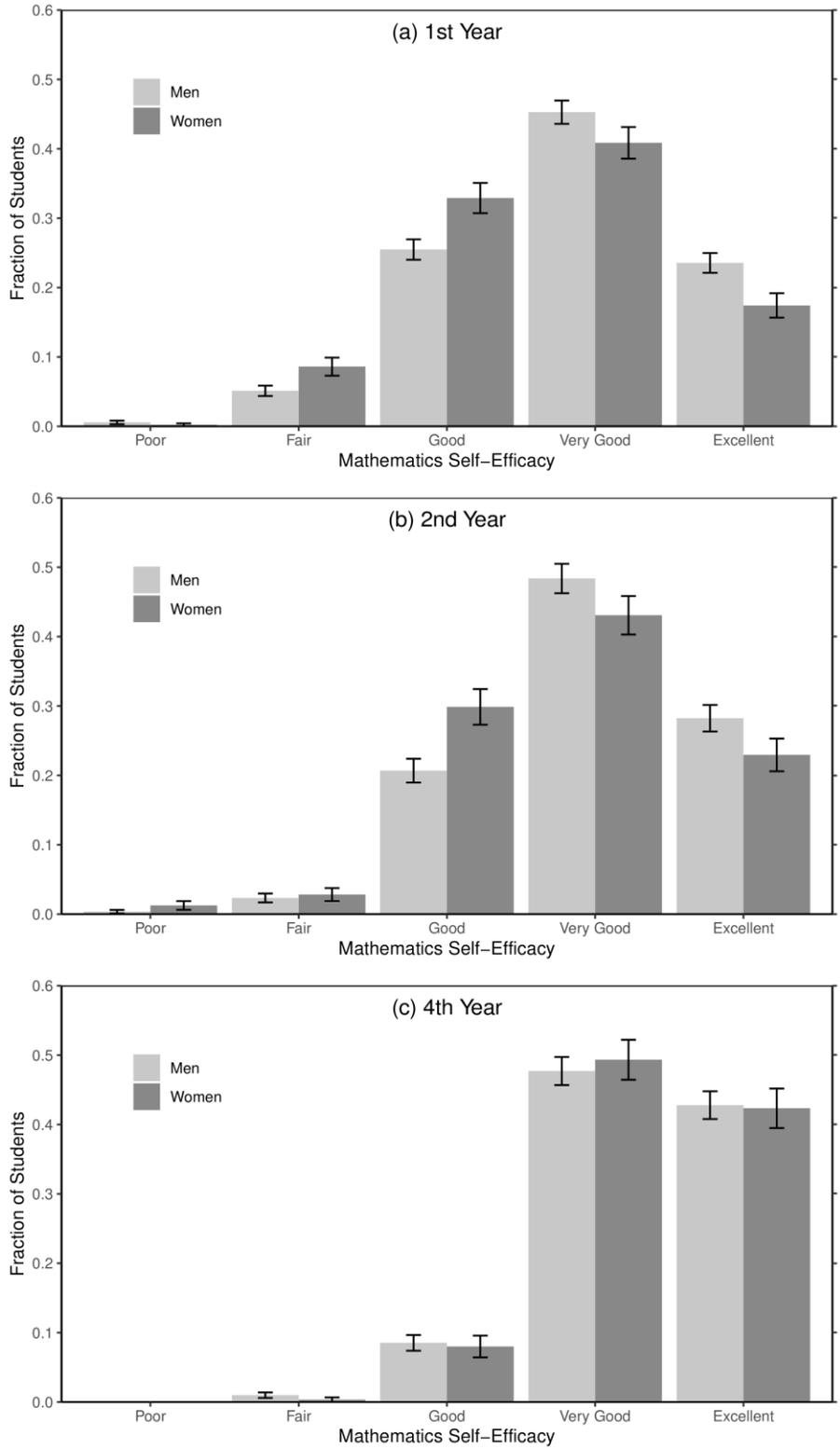

Figure 6: Distribution of responses to the mathematics self-efficacy prompts the surveys taken by students in their (a) 1st year, (b) 2nd year, and (c) 3rd year. The fraction of men and women, respectively, who answered each of the five options are plotted along with the standard error.

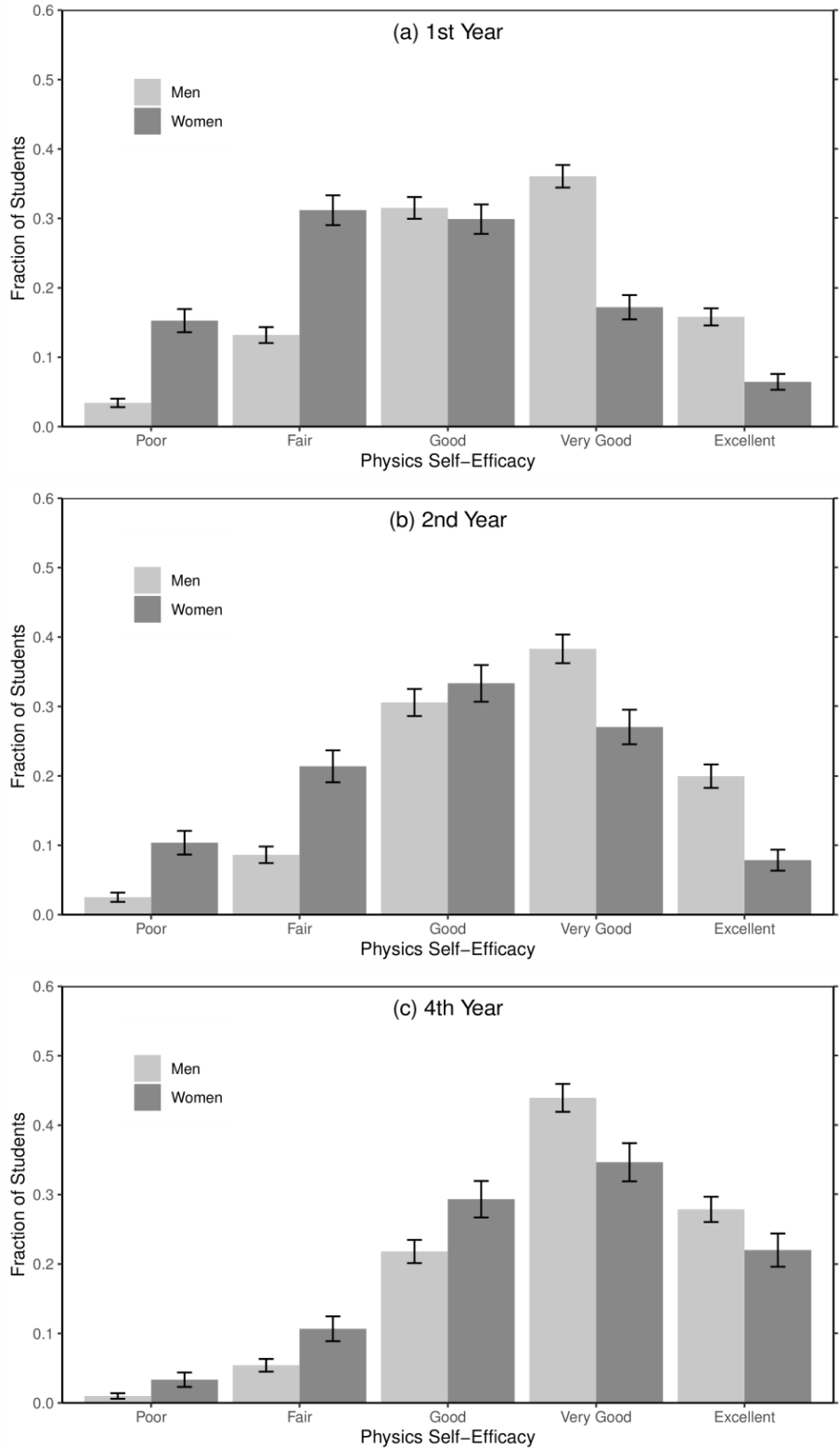

Figure 7: Distribution of responses to the physics self-efficacy prompts the surveys taken by students in their (a) 1st year, (b) 2nd year, and (c) 3rd year. The fraction of men and women, respectively, who answered each of the five options are plotted along with the standard error.

# APPENDIX B: Self-Efficacy Over Time by Major

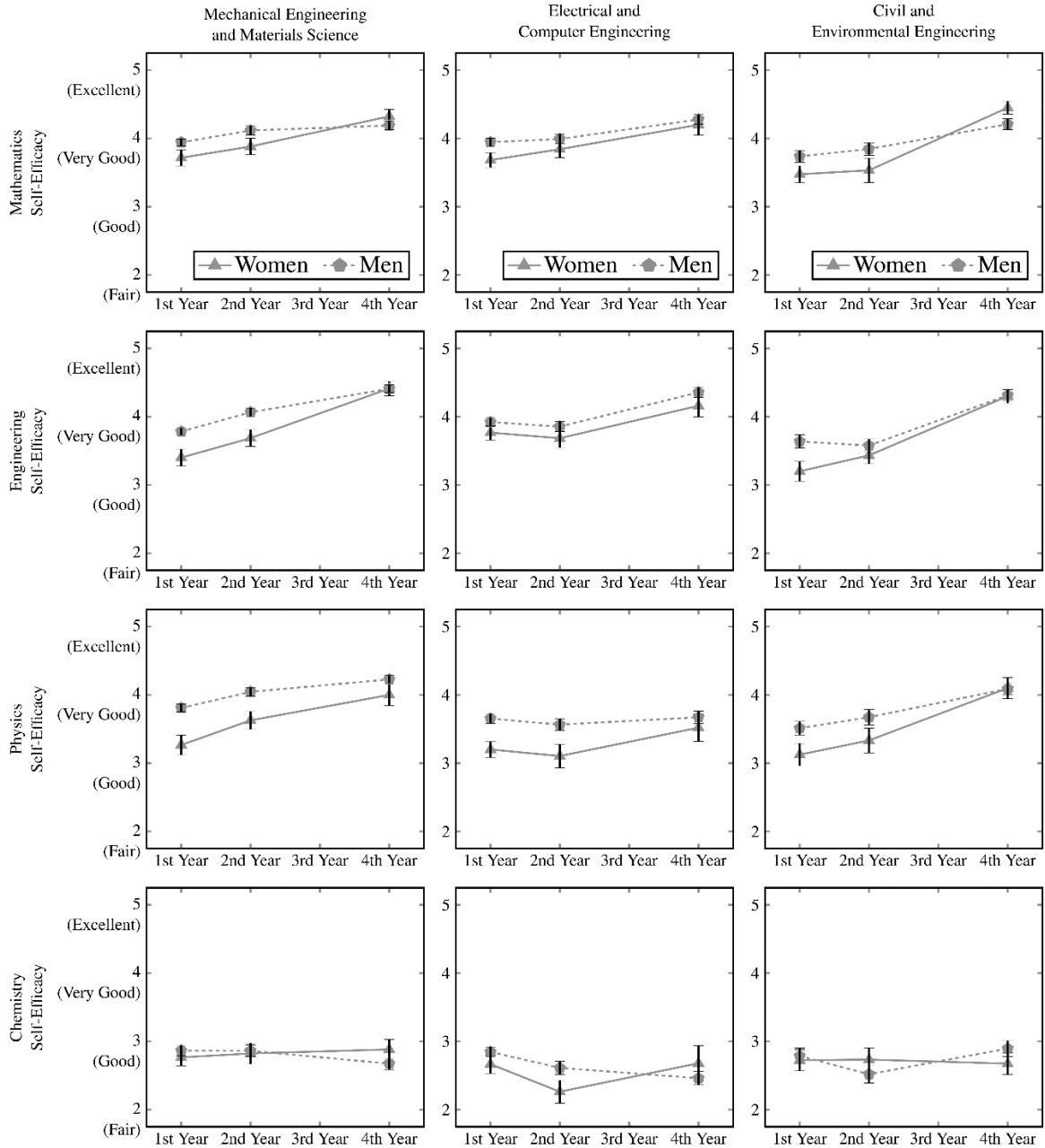

Figure 8: As in Figure 3, the mean self-efficacy scores of engineering students at the end of their first, second, and fourth years in each of the foundational subjects in engineering are plotted along with their standard error. These results are now plotted separately for students in each of the six majors (the three majors with the lowest percentage of women in this figure and the three with the highest percentages of women in Figure 6). Self-efficacy was measured on a Likert scale from 1 to 5. Each column contains the graphs for the different majors while each row contains the graphs for self-efficacy in the different foundational subjects.

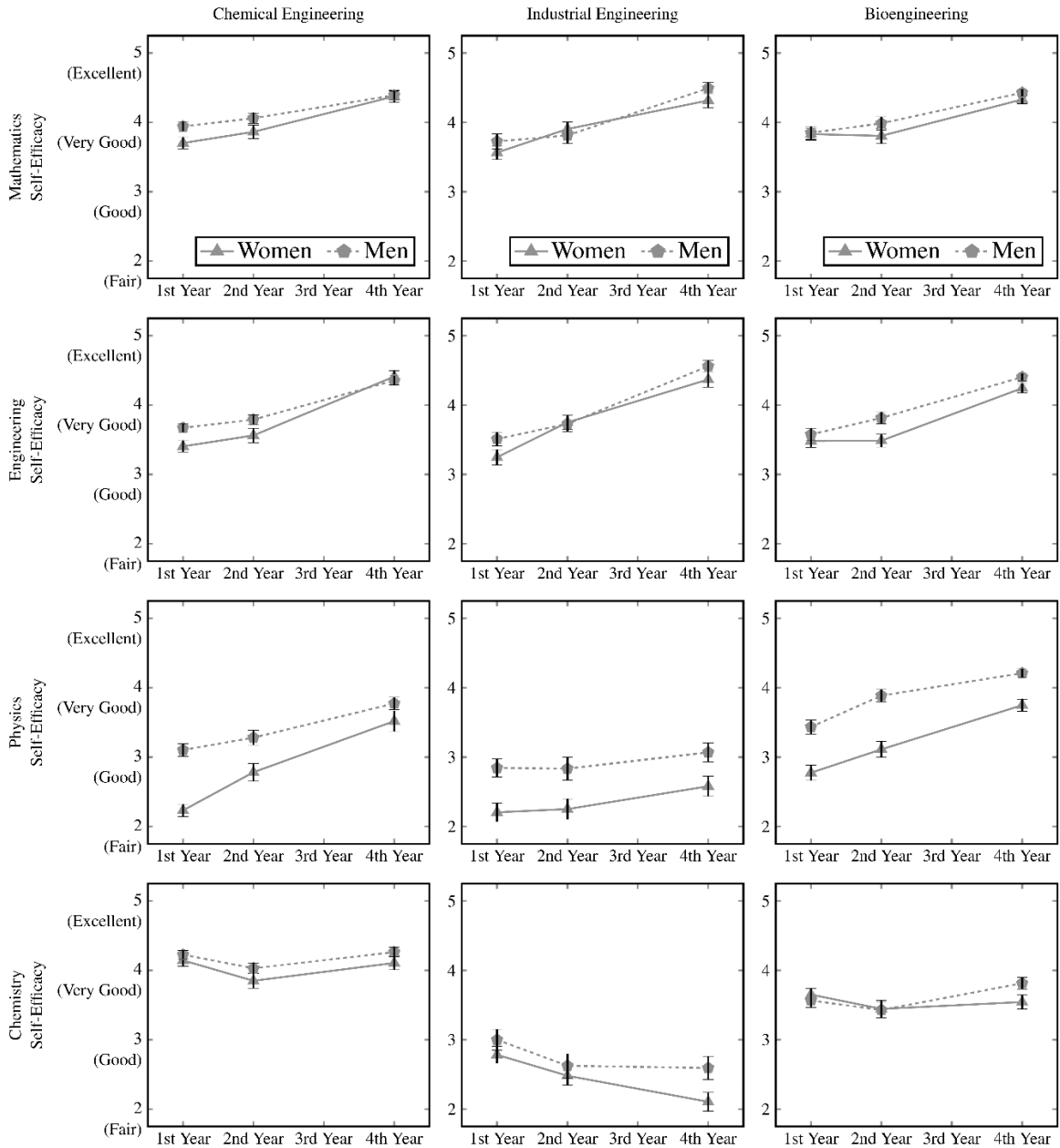

Figure 9: As in Figure 3, the mean self-efficacy scores of engineering students at the end of their first, second, and fourth years in each of the foundational subjects in engineering are plotted along with their standard error. These results are now plotted separately for students in each of the six majors (the three majors with the lowest percentage of women in Figure 5 and the three with the highest percentages of women in this figure). Self-efficacy was measured on a Likert scale from 1 to 5. Each column contains the graphs for the different majors while each row contains the graphs for self-efficacy in the different foundational subjects.